# Giving Mechanical Engineers Intelligent Tools: A Project-Based AI Education Curriculum in Thermal Engineering

Changgen Li, Han Hu, *Member, IEEE*, Christy Dunlap, Nathaniel House, and Jonathan Wai

***Abstract*—Mechanical engineering (ME) requires a broad knowledge base across several disciplines; however, ME students often have insufficient training in electrical and computer engineering, complex challenges in traditional thermal system modeling, and endure heavy course loads with limited class hours. To help address these challenges, this paper proposes a new curriculum that integrates artificial intelligence (AI) into ME at the University of Arkansas (UARK), with a particular emphasis on thermal problems and their interplay with electrical and computer engineering. The curriculum has introductory, application, and advanced levels, covering core and optional AI projects. Key goals are to enhance students' understanding of AI models, ability to tackle engineering tasks, and teach multidisciplinary communication skills. This curriculum offers educators and researchers valuable insights into courses that can enhance students' practical skills and creativity. This curriculum, including the syllabus, data, and codes, is available to the public in open-access repositories.**

## I. INTRODUCTION

Mechanical engineering (ME) encompasses the design, manufacturing, operations, and maintenance of various mechanical systems and equipment [1], [2]. These tasks require MEs to acquire expertise in a wide variety of knowledge domains across science, technology, engineering, and mathematics (STEM) [3]. Fig. 1 illustrates the ideal knowledge structures of an ME education based on the current course sequence at the University of Arkansas (UARK). These multidisciplinary concepts and skill sets can enable students to comprehend the principles, structure, functionality, and performance of an electro-mechanical-hydraulic integrated system. This can also help students collaborate and communicate more effectively with experts in other fields such as electrical engineering, materials science, computer science, and beyond. Oftentimes, without having significant knowledge and grounding in adjacent disciplines, it can be hard even for highly technical ME experts to be able to effectively communicate with other experts on their product or engineering teams in the workplace [4]. The ME authors on this paper specialize in thermal engineering, so the focus on the science of learning and education will mostly be in this subspecialty and from the higher education research and training perspective.

As a specialized branch of ME, thermal engineering focuses on the study, design, and application of systems and processes involving engines [5], HVAC (heating, ventilation, and air conditioning) systems [6], power plants [7], refrigeration systems [8], and thermal management in electronic devices [9]. Artificial Intelligence (AI) [10] is a natural complement to the learning and education of engineers in general, MEs, and thermal engineering specifically. Despite the current strengths of ME education, there remain a few opportunities to further improve the curriculum for ME students, especially by integrating AI into their learning and education.

1) ME students have a limited background in AI: Recent studies consistently emphasize the need to systematically integrate computational skills and interdisciplinary competencies into engineering education to enhance students' ability to address complex, cross-domain engineering problems and strengthen their professional competitiveness [11], [12], [13], [14]. However, programs such as UARK's BS in ME still focus heavily on dynamics and thermofluids, offering limited exposure to electrical engineering and programming. This gap can hinder graduates' ability to collaborate effectively with experts from other disciplines.

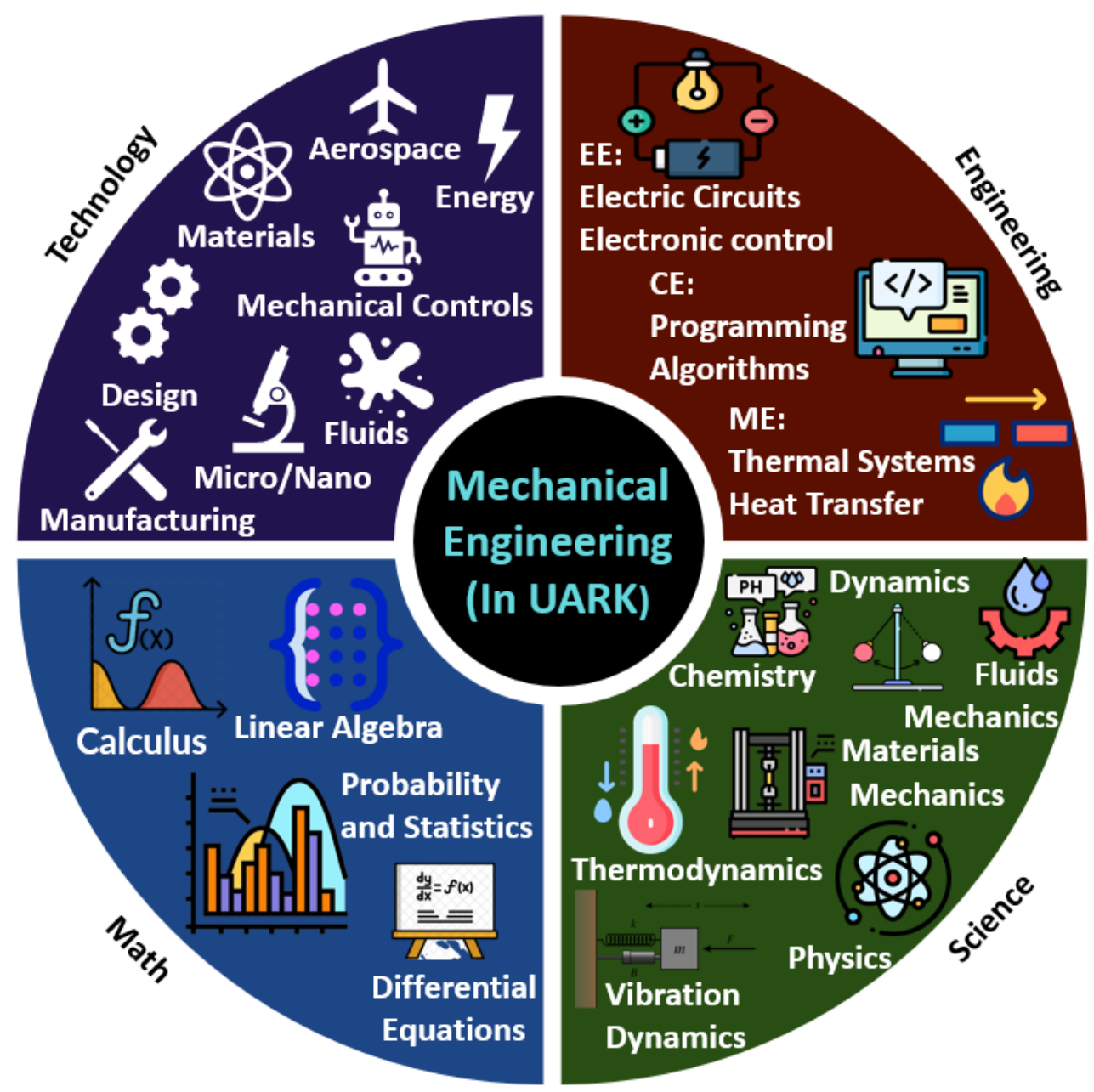

Fig. 1. Multidisciplinary knowledge structure of the current mechanical engineering curriculum for undergraduate students at the University of Arkansas. These knowledges contain science, technology, engineering, and math (STEM). They are diverse and time-intensive, yet they offer limited content that intersects Electrical Engineering (EE), Computer Engineering (CE), and Mechanical Engineering (ME) disciplines. Consequently, the challenge of enabling students to efficiently acquire interdisciplinary knowledge and enhance their problem-solving skills under constrained class hours and workloads has motivated the development of an AI-integrated ME curriculum.

2) Current challenges in traditional thermal systems modeling: Traditional thermal system modeling often relies on physical laws and empirical equations such as the law of conservation of energy, the heat conduction equation, and the Navier-Stokes equation in an attempt to accurately describe physical processes [15]. These physical methods typically have difficulty capturing interactions and nonlinear behaviors in a complex system, particularly when facing multiphysics and multiscale problems [16]. Moreover, the abstract and complex mathematics related to thermodynamic theories can be quite challenging for students, making thermal system modeling a daunting task [17].

3) The heavy course load and limited class hours: ME curricula are typically highly structured, comprising numerous required courses in mathematics, sciences, and core engineering topics, as illustrated in Fig.1. Such rigidity often results in a heavier workload than many other majors, leaving students with limited flexibility to explore emerging areas of interest [18]. Research indicates that adding additional standalone courses to already dense programs can increase academic stress and potentially extend time-to-graduation [19]. Furthermore, the limited class hours constrain the range of foundational topics that can be addressed [20]. For example, many ME programs allocate minimal instruction time to electrical engineering or computer science, hindering students' interdisciplinary readiness. Therefore, requiring ME students to take long-duration AI courses offered by computer science departments would likely further increase their academic workload.

4) Current AI education, typically machine learning (ML) or deep learning (DL), occurs largely in computer science departments. As a result, these AI courses are more oriented towards computer science tasks like image recognition and natural language processing. They prioritize algorithms, programming, and data processing while offering limited attention to specific issues and practical applications that would be useful in other disciplines, such as engineering [21]. For ME students, these AI courses may lack direct relevance and practicality, depriving future MEs of sufficient insight into leveraging AI to better understand physical principles and mechanisms, or to solve engineering problems using AI in useful ways [22].

Therefore, to overcome these challenges, it is necessary to design a curriculum for ME students that integrates AI with the existing thermal engineering topics. This pathway should ideally enhance students' knowledge in electrical and computer engineering with a reasonable number of class hours, while also hopefully simplifying the complexity of thermal system modeling.

On the other hand, despite the proven success of AI-assisted research and development, the current implementation heavily relies on the AI algorithms developed and matured in other areas such as natural language processing and computer vision. The development of more sophisticated ML and DL models for thermal engineering problems requires researchers trained in both data science and engineering disciplines. To this end, the concepts and skills related to AI need to be synergistically incorporated into the thermal engineering curriculum. As AI has become more prevalent, many researchers have started to introduce AI into their programs. For example, Muqri et al. [23] discussed ML education using Python, Octave, and MATLAB. Their work provides a quick start for beginners using miscellaneous programming tools. Different from Muqri et al.'s work, which requires a background in a specific programming language, Zhang [24] introduced a one-credit-hour ML course taught using Excel. This course can effectively reduce the initial programming barrier for students. Besides the above-mentioned works that focus on teaching students AI algorithm coding, some researchers have also attempted to embed AI into specific engineering fields. Kim et al. [25] explored the integration of ML models into a curriculum focused on vibration signal analysis, highlighting the benefits of incorporating ML into ME education. Their findings indicate that students, regardless of whether they participated in the ML course, exhibited strong enthusiasm for ML and recognized its critical importance to their professional development. Furthermore, their research demonstrates that hands-on practice significantly enhances students' confidence in applying ML techniques and can foster greater interest in mechanical vibration. Chalacheva reported the development of an introductory course on AI and ML in biomedical engineering [26]. This course serves as an effective bridge, enabling students in biomedical engineering to transition or expand their expertise from the biological sciences to engineering. In addition, AI has also been introduced into rudimentary STEM courses as extra-credit assignments or course modules, for example, the AI-related course modules for engineering, astronomy, sociology, and anthropology by the NSF-funded data science infused into the undergraduate STEM curriculum (DIFUSE) project at Dartmouth College [27]. According to Petra Bonfert-Taylor, Professor at the Dartmouth School of Engineering and Principal Investigator of the DIFUSE project, the initiative has significantly improved students' preparedness for entering the job market [28]. At UARK, a data science degree program has been developed and hosted in the College of Engineering. Many engineering students who are passionate about AI are also enrolled in AI courses offered by the computer science department. However, to the best of our knowledge, there is currently no publicly accessible project-based AI course specifically for thermal engineering. Nor is there an AI course for mechanical engineering that features a wealth of student academic publications.

The contributions of this paper are as follows:

1) We introduce a project-based AI curriculum on thermal engineering specifically designed for ME students. This curriculum is open access for all, including STEM instructors, education researchers, and those in other disciplines. This curriculum may also serve as a reference for other educators or researchers to improve education effectiveness or quality, and to draw inspiration from, use, or build on it.

2) We illustrate what the authors believe to be the power of project-based AI education for MEs by emphasizing hands-on problem-solving instead of course examinations. This curriculum inspired students to design innovative projects that may span different engineering categories. Perhaps the curriculum and ideas presented here might be beneficial for research, initiating collaborations and learning between the fields of STEM and education.

The rest of this manuscript is organized as follows. Section II provides the impact of AI-integrated education and a general background in thermal engineering. Next, the importance of integrating AI into thermal engineering education is explained. Section III presents the goals, hierarchy, and project details of this new curriculum. Section IV illustrates the student performance assessment, analysis, and examples of outcomes. The conclusion is given in Section V.

## II. Related Background of AI for Engineering

### *A. The impact of AI-Integrated curriculum*

Integrating AI into the curricula of various disciplines has likely influenced student learning experiences and outcomes in multiple ways. Park et al. [29] and Walter [30] find that AI-integrated science lessons help students develop critical thinking skills and AI literacy, which are essential for navigating the complexities of modern technology. Students learn not only how to use AI tools but also how to evaluate their outputs critically and understand their limitations and ethical implications. Park et al. [29] also note that AI curricula often involve project-based learning that applies AI to real-world problems, fostering interdisciplinary collaboration. These projects demonstrate how students can integrate AI with other disciplines like STEM studies, enhancing their practical problem-solving skills. Preparing students with these skills can ensure they remain competitive in a rapidly evolving job market [31]. Integrating AI into STEM curricula also allows educators to measure their impact on student performance and career readiness. It provides insights into how well students are prepared to apply AI in their future professional roles, ensuring that the curriculum meets evolving industry demands [30].

### *B. Thermal engineering*

The primary objective in thermal engineering is to design systems that efficiently manage and utilize heat, thereby optimizing energy use and reducing safety risk. Take data centers as an example. According to statistics [32], the cooling systems in data centers typically consume around 45% of the total energy, while IT equipment, as the core components of data centers, account for only about 40%. Thus, designing high-performance thermal management systems is crucial for controlling operational costs and optimizing data center efficiency. On the other hand, a thermal crisis known as critical heat flux (CHF) during two-phase (like boiling) cooling can damage cooling equipment. This is because when CHF is reached, a thin film may form on the heat transfer surface, leading to a sharp temperature increase on the surface, potentially burning the cooled object. Traditional methods regulate the cooling system's flow by monitoring temperature or pressure to prevent crises. However, traditional methods only rely on some specific physical quantities but ignore the interplay within these quantities. Additionally, traditional methods cannot effectively utilize the equipment's design parameters, operational parameters, and historical monitoring data. Therefore, they often exhibit poor accuracy and high hardware costs. Different from the traditional methods, AI-implemented models possess superior feature extraction and information fusion capabilities. Consequently, they are not only suitable for monitoring but also for diagnosis and prediction. Based on this, the authors argue that integrating AI into thermal engineering can yield several advantages: i) Achieving more precise and robust predictions of heat flux based on various design and operational parameters, such as heater geometry [33], pressure [34], and flow rate [35]. ii) Integration of sensing and metrology. AI-based (or data-driven) modeling can fuse a wide array of sensing approaches ranging from two-dimensional signals like optical images and thermographs to one-dimensional signals like acoustics, temperature, and pressure [36], [37], [38], [39], [40]. iii) Extracting physics descriptors to probe the underlying heat transport mechanisms. AI can leverage the understanding of transport mechanisms during boiling by identifying physical descriptors of boiling states based on sequences of optical images[41], correlating thermal parameters like heat flux with design and operational parameters such as surface tension, enthalpy of vaporization, and thermal conductivity [42].

Therefore, integrating AI into the thermal engineering curriculum is necessary for students to gain valuable interdisciplinary skills and knowledge that are crucial for developing advanced thermal management systems, optimizing energy use, and improving overall system efficiency and safety. This curriculum can also provide an example for educational researchers to design effective learning experiences and assessments, perhaps helping develop more comprehensive educational models that integrate AI with STEM principles.

## III. An AI Curriculum for Thermal Engineering

### *A. Curriculum goals*

At UARK, the authors developed a course named "Machine Learning for Mechanical Engineers (MLME)" that follows a project-based learning strategy and integrates AI into cutting-edge development in engineering. Distinct from AI courses in computer science or data science, the MLME course focuses on the proper selection and implementation of AI algorithms for enhanced problem-solving in thermal engineering. The overarching goal of the MLME course is to equip mechanical engineers with ML skills and deepen the integration of data science into the ME curriculum. Students are provided with concrete and specific engineering problems with experimental data. The projects, presentations, and in-class peer review practice are designed to foster students' professional skills following the National Association of Colleges and Employers (NACE) competencies [43], including critical thinking, communication, teamwork, technology, leadership, and professionalism. Students completing this course are expected to be able to develop, train, and test ML models using Python (the TensorFlow framework) and MATLAB, develop ML models for image classification and clustering, perform data dimensionality reduction for physics extraction, analyze images from experiments and simulations to predict physical quantities, adapt trained ML models for new applications, analyze time series for classification and regression, and develop surrogate models for computationally expensive numerical simulations.

*B. Curriculum hierarchy*

The MLME course is a project-based course designed according to the principles of scaffolded instruction [44], i.e., gradually reducing guidance as students gain mastery, which is represented as a pyramid in Fig. 2. At the base of the pyramid lies curiosity and self-motivation. It serves as the foundational attitude driving students to explore and engage deeply with the study material. In this course, student curiosity and self-motivation are intentionally cultivated both in the general introduction and at the outset of each project module. Specifically, in the general introduction phase, students first understand the development of scientific theories in thermal engineering modeling and their inherent limitations. They then explore the emergence of AI technologies and how these can address the shortcomings of thermal system modeling based solely on fundamental theories and empirical formulas. To further inspire engagement, real-world cases are presented comparing salary levels, career path diversity, and professional ceilings between traditional mechanical engineering graduates and those with AI expertise. During each project module, students are first guided to recognize the challenges of solving the targeted thermal problem using conventional physics-based methods. The discussion then transitions to AI-based solutions, highlighting their advantages through comparison. Finally, simple accessible examples are provided to demonstrate that implementing AI methods can be easy and straightforward, thereby reducing psychological barriers to learning and encouraging active participation. A well-designed introductory course plays a crucial role in enhancing learning outcomes by motivating students to actively engage, thereby improving academic achievement and transforming learning into an interesting, dynamic experience rather than a passive, dependency-driven process [45], [46].

The second level from the base of the pyramid is foundational knowledge. Students need to grasp some foundational STEM knowledge before understanding more complex concepts [47]. This level will be given in each project. Basic matrix operations and fundamental programming syntax in MATLAB and Python are introduced to students in the first project. In subsequent projects, foundational knowledge is provided in a just-in-time manner aligned with the problem at hand. For example, when advancing to more sophisticated programming, students are shown how to implement algorithms not through manual matrix operations alone, but by leveraging pre-built packages developed by computational scientists. This enables students to focus more on designing algorithmic architectures and programming logic to address thermal engineering problems.

Building upon foundational knowledge, the mathematical and statistical principles of ML techniques, such as polynomial fitting, Gaussian statistics, and hypothesis testing, are introduced at an introductory level, providing students with a solid grounding in AI tools. On the other hand, the thermal problem of each project is also introduced here. This component of the course guides students in abstracting an engineering problem into a corresponding mathematical formulation and categorizing it into AI-relevant problem types, such as classification, regression, or time-series forecasting. Based on the resulting thermal problem model, students can then accurately identify and select the most appropriate AI algorithms to address the problem. It should be noted that the “Introduction level” in the pyramid refers to the stage of teaching students the core theories and techniques, which is distinct from the “Introduction” at the beginning of the curriculum that provides a general overview of the course.

Subsequently, the course progresses to the application level, focusing on exploring the potential applications of the knowledge introduced at the introductory level while expanding on theories. The objective of this level is to practice ML algorithms using publicly available datasets in the computer field, thus consolidating foundational knowledge while enhancing practical skills. Additionally, theoretical aspects of signal processing, such as the Fourier transform and time series analysis, are also discussed. At the advanced level, thermal theory topics such as heat flux, boiling processes, bubble dynamics, and thermal experiment datasets are covered in detail. Based on these thermal experiments, students are required to complete four projects integrating ML into thermal engineering, and possibly even develop novel ML models for thermal engineering problems.

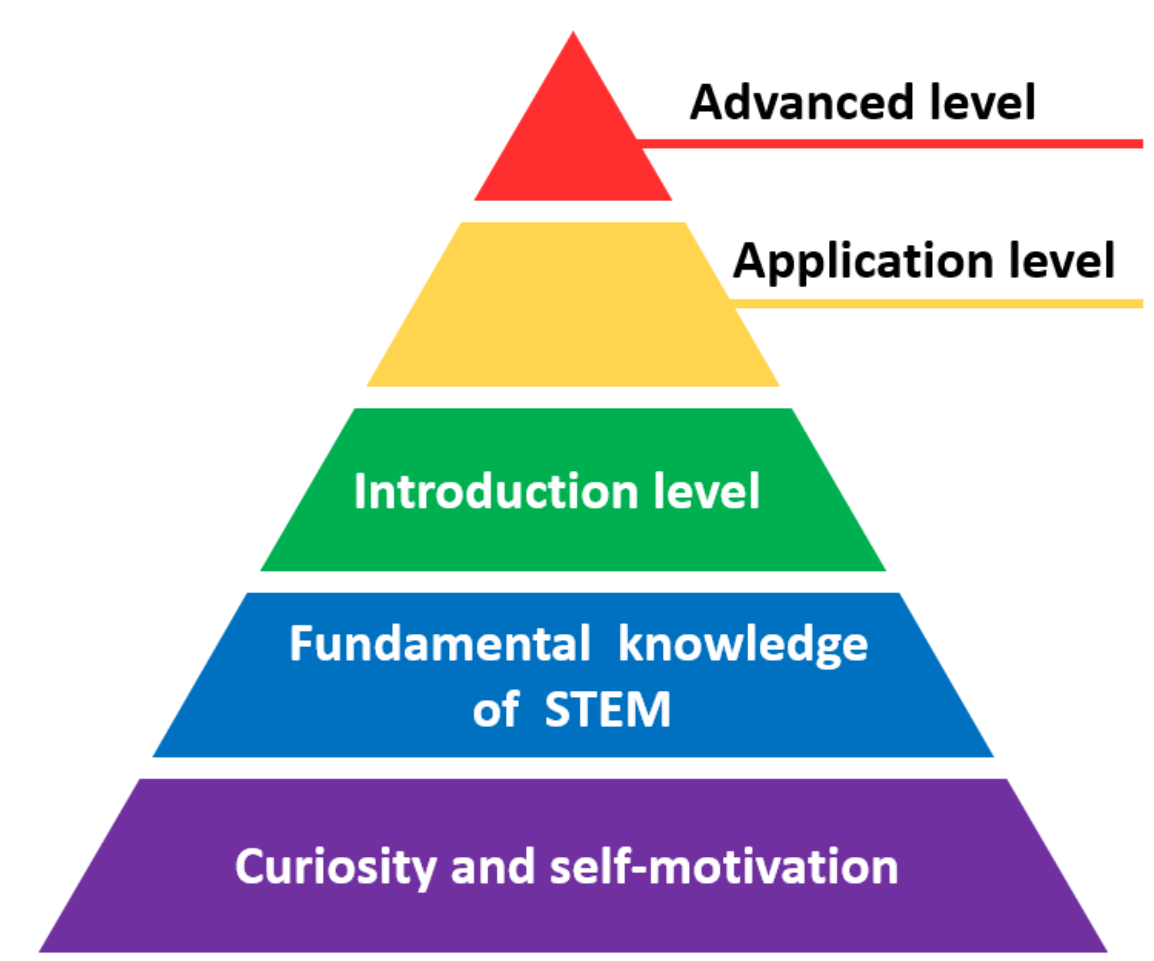

Fig. 2. Hierarchy of the AI-integrated engineering curriculum. This "Machine Learning for Mechanical Engineers (MLME)" course is structured to guide students through multiple levels of learning, represented by a pyramid. At the base of the pyramid is "Curiosity and self-motivation," which serves as the foundational attitude driving students to explore and engage deeply with the material. The next level, "Fundamental knowledge of STEM," ensures that students have a strong grounding in the essential principles of science, technology, engineering, and mathematics. Moving up, the "Introduction level" introduces the basic concepts and techniques of machine learning, tailored specifically for mechanical engineering applications. The "Application level" then focuses on practical implementation, allowing students to apply their knowledge to real-world problems and projects. Finally, the "Advanced level" provides an in-depth exploration of complex machine learning methodologies, preparing students for cutting-edge research and professional excellence in the field of mechanical engineering.

*C. Curriculum entry requirements*

The preferred entry requirements of MLME include three aspects, i.e., mathematics, programming, and thermal. First, basic mathematics: one semester of calculus, linear algebra fundamentals, and basic probability/statistics. These are essential for understanding machine learning algorithms and developing mathematical models of engineering problems. Second, introductory programming skills: familiarity with MATLAB or Python syntax, data manipulation, and basic plotting. This ensures students can implement and test algorithms without facing steep programming barriers. Third, core mechanical engineering knowledge: completion of at least one thermal sciences course (e.g., thermodynamics or heat transfer) to enable understanding of domain-specific ML applications in thermal engineering. However, high-school level mathematical and physical knowledge is the minimal prerequisite since the instructional design follows a scaffolded approach, starting from the most fundamental concepts and progressively introducing more complex ideas and applications. Essential mathematical and programming skills are taught or reinforced within the course itself, ensuring that students without prior exposure can still engage meaningfully with the content.

*D. Curriculum structure and project detail*

The course structure is organized into a timeline format as shown in Fig. 4, totaling 39.1 hours, and covers various topics such as linear regression, multilayer perceptron (MLP), convolutional neural networks (CNN), and time series analysis. The distribution of course time is strategically allocated to emphasize different topics. The course content is closely integrated with four assignment projects. Here, we detail the four projects and their required background knowledge. Boiling, as shown in Fig. 3, is a complex phase-change phenomenon during heat transfer, where a liquid transforms into a gas upon contacting a heated surface. This process involves several crucial concepts, such as heat flux and boiling regimes. Heat flux is the rate of heat energy transfer per unit area and plays a central role in boiling, as it determines the rate and intensity of the heat transfer (bottom panel of Fig. 3). Boiling regimes refer to the different stages within the boiling process, each characterized by specific features and heat transfer mechanisms. For instance, the process begins with convective heat transfer where there is no bubble generation, and the heat is directly transmitted from the heating surface to the liquid. As the boiling intensity increases, nucleate boiling starts, where bubbles form at discrete points on the heated surface (bottom panel of Fig. 3). As nucleate boiling intensifies, the morphology of bubbles changes until they reach the critical heat flux (the second dashed line from the left, bolded and marked in red). At this point, the rate of liquid evaporation undergoes significant changes. Beyond the critical heat flux point, the system enters the film boiling regime. In this stage, the heated surface is covered by a layer of gas, which reduces direct contact of the heater surface with the liquid and decreases the efficiency of heat transfer and can even cause device failure. Therefore, it is necessary to continuously monitor the heat flux and boiling status. Correspondingly, the four assignment projects are developed to achieve online boiling monitoring, diagnosis, and prediction goals.

Project 1 focuses on heat flux analysis (see Fig. 4). This project belongs to a monitoring task. In pool boiling experiments, the boiling heat flux can be estimated as the supplied power divided by the heater surface. However, this estimation will not be very accurate due to heat loss and other non-ideal conditions in experiments, especially for thin-film heaters with relatively low thermal conductivities (e.g., Indium Tin Oxide heaters). Conventionally, finite-element simulations are used to evaluate heat loss to validate or correct the experimental assumptions. ML provides another perspective for tackling this issue. The heat loss and other non-ideal conditions can be captured and accounted for by the hidden layers of neural networks. In this project, students need to set up and train an MLP and a Gaussian process regression (GPR) model to predict the heat flux based on the temperature data, report the training curves (training/validation accuracy/loss vs. epoch), the training time (time/epoch, time until the best model), and then circumvent the effects of overfitting using k-fold cross-validation (e.g., using 100 folding).

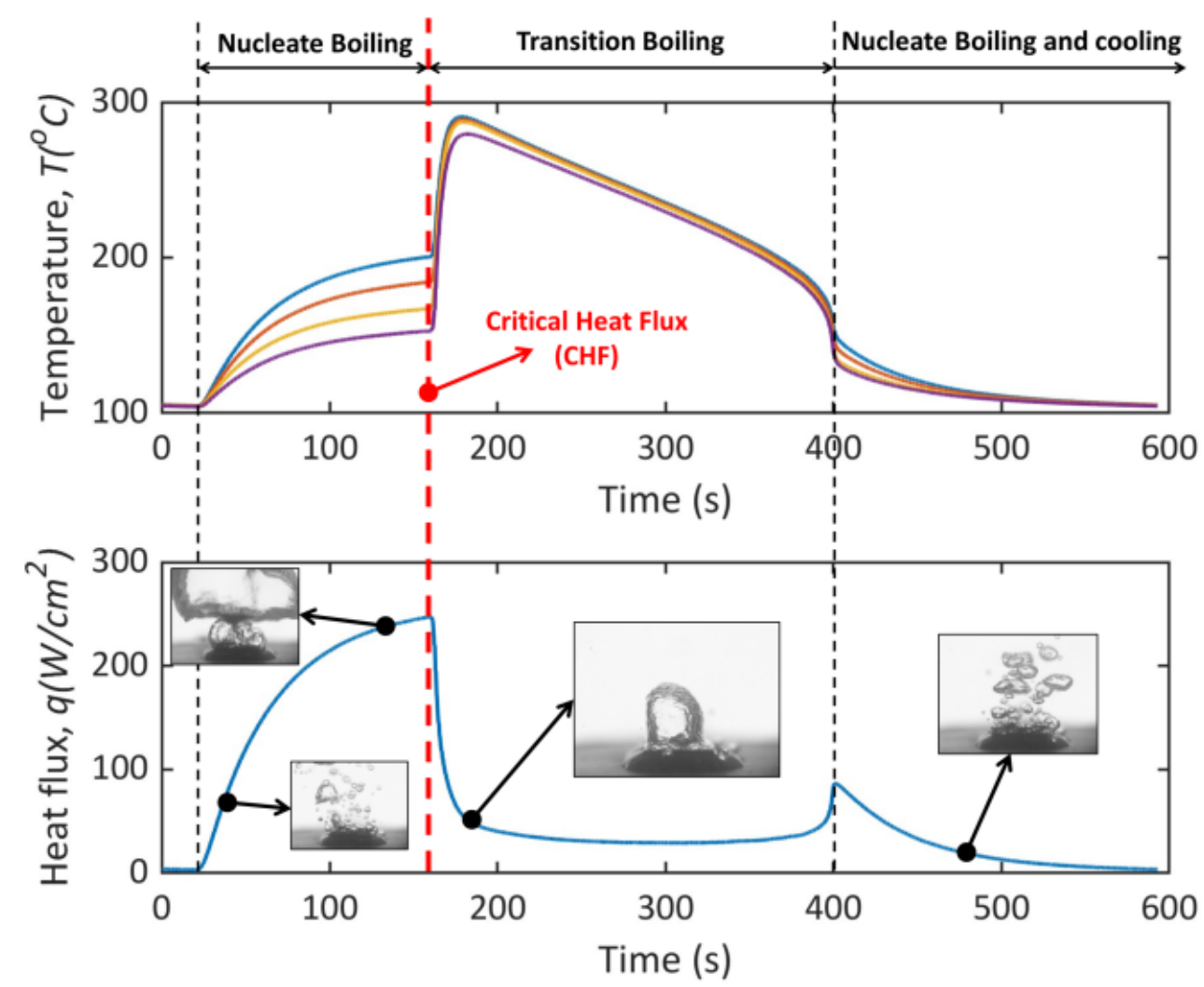

Fig. 3. Boiling features. Top panel: Temperature changes over time during boiling. Bottom panel: Heat flux changes over time and the bubble morphology at different boiling regimes. Initially, the system begins with convective heat transfer, where no bubbles are generated, and heat is directly transmitted from a heating surface to the liquid. As the heat intensity increases, nucleate boiling starts, with bubbles forming at discrete points on the heated surface. As the nucleate boiling intensifies, the morphology of the bubbles changes until they reach the critical heat flux (CHF) point, indicated by the red dashed line. At the CHF, the rate of liquid evaporation undergoes significant changes. Beyond this point, the system enters the film boiling regime, where the heated surface is covered by a vapor layer, reducing direct contact with the liquid, decreasing heat transfer efficiency, and potentially causing device failure. Therefore, it is necessary to continuously monitor the heat flux and boiling status. Correspondingly, the four assignment projects are designed to achieve the goals of boiling monitoring, diagnosis, and prediction.

Project 2 addresses supervised boiling regime recognition. This project is a two-class diagnosis task. Boiling has been widely implemented in the thermal management of high-power-density systems, e.g., nuclear reactors, power electronics, and jet engines, among others. The CHF condition is a practical limit of pool boiling heat transfer. When CHF is triggered, the heater surface temperature ramps up rapidly (~150°C/min), leading to detrimental device failures. There is an increasing research interest in predicting CHF based on boiling images. Two class images, "pre-CHF" and "post-CHF" that reflect pool boiling status before and after CHF is triggered, are provided. The students need to split the data set into training, validation, and testing. This can be done before training with a separate package or directly in the code during training by setting up and training a model to classify the pre-CHF and post-CHF images using CNNs or pre-trained CNNs. Reporting of the training curves (training/validation accuracy/loss vs. epoch) and the training time (time/epoch, time until the best model) is required. Moreover, model tests should be carried out using the reserved test data, and should report the confusion matrix, accuracy, precision, recall, F1 score, the receiver operating characteristic (ROC), and area under the curve (AUC).

Project 3 involves unsupervised classification. This project also belongs to a two-class diagnosis task but in an unsupervised way. The students are required to run a dimensionality reduction and clustering analysis of the boiling image dataset for unsupervised boiling regime recognition. In this project, they need to run single value decomposition (SVD) or principal component analysis (PCA) of the images and plot the percentage explained variance vs. the number of principal components (PC). Then, they are asked to pick a representative image, run PCA, and plot the reconstructed images using a different number of PCs. They then must calculate the error of the reconstructed images relative to the original image and plot the error as a function of the number of PCs. Finally, students are asked to run a clustering analysis of boiling images using the PCs and evaluate the results of clustering.

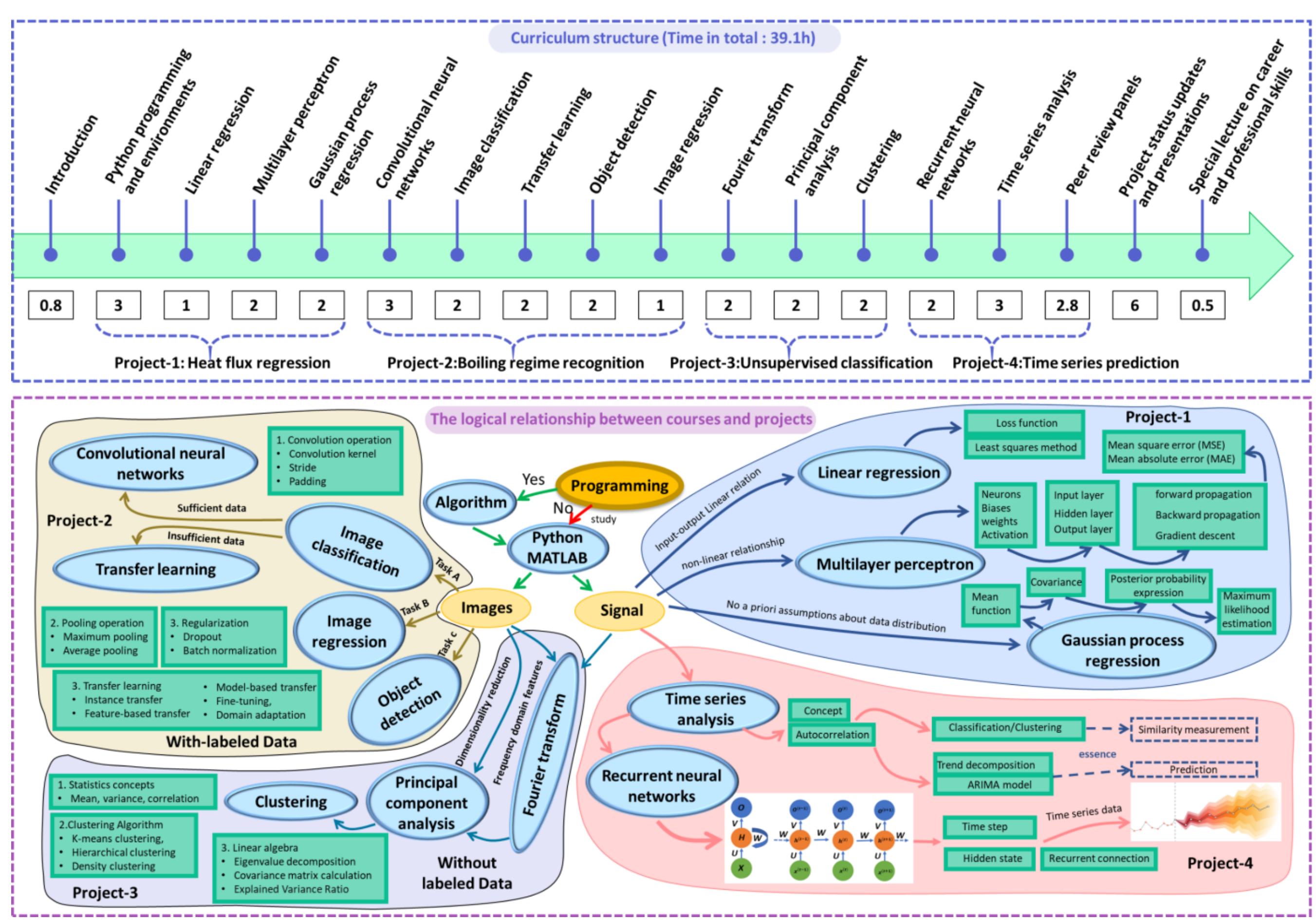

Fig. 4. Visual structure of the Machine Learning for Mechanical Engineers (MLME) curriculum. This project-based curriculum is meticulously structured to span a total of 39.1 hours, integrating theory and practical applications through four comprehensive projects. The curriculum begins with foundational programming skills in Python and MATLAB, focusing on algorithms for image and signal processing. Project 1 introduces linear regression, multilayer perceptron, and Gaussian process regression, building a strong base in predictive modeling and error analysis. Project 2 dives into convolutional neural networks and transfer learning for image classification, regression, and object detection, emphasizing both sufficient and insufficient data scenarios. Project 3 explores clustering algorithms and principal component analysis, addressing both labeled and unlabeled data. Finally, Project 4 covers time series analysis and recurrent neural networks, enabling students to handle temporal data and perform trend analysis and prediction. This curriculum ensures a logical progression of topics, fostering a deep understanding of machine learning principles and their application in mechanical engineering.

Project 4 concentrates on time series prediction. In this project, students need to develop a recurrent neural network (RNN) to forecast the vapor fraction (a parameter of bubbles in an image or a frame of a video) of future frames based on past frames. The vapor fraction represents the proportion of vapor relative to the total volume or mass of the mixture. It is a dimensionless value that ranges from 0 (completely liquid, with no vapor) to 1 (completely vapor, with no liquid). This parameter is critical for understanding the behavior of mixtures in various engineering and scientific applications, enabling precise control and optimization of processes involving phase changes. The students need to plot the model-prediction vs. the ground true and vary the input and output sequence lengths to evaluate their effect on the error of the model predictions.

Additionally, three extra projects are provided as options for the final exam project, and students are also encouraged to propose their own projects. These projects are focused on using the ML algorithms practiced in the assigned Projects 1 through 4 in tandem to solve more complicated problems. Extra Project-1 is named PCA-MLP in Tandem. In this project, students need to redo the image classification problem in Project-2 using PCA-MLP and run SVD or PCA to obtain the PCs of the images. Then, they need to feed the PCs to an MLP neural network to classify the regime of the boiling images. Extra Project-2 is named Image Regression. This project is also correlated with the original Project-2. For the Image Regression project, students need to build a CNN model to predict the vapor fraction of the images and compare the model prediction against the true data. Extra Project-3 is named High-dimensional seq2seq Learning. In this project, the students need to build an RNN model including long short-term memory (LSTM) or Bidirectional-LSTM. They then need to take the PCs of an image sequence as input to forecast the PCs of future frames.

## IV. Student performance assessment, analysis, and Examples of Outcomes

### *A. Student performance assessment*

All the courses are administered by the instructor, who takes on the responsibility of project preparation. The course grade, totaling 100 points, is equally divided between four assignment projects (50 points) and a final project (50 points). Each assignment, worth 12.5 points, requires a written report and corresponding code, evaluated based on machine learning model development (40%), successful implementation (30%), and report/presentation quality (30%). For the final project, students choose a topic from the provided options, encouraged to align with their research interests and to think creatively. The final project includes multiple deliverables: a kick-off presentation (2.5 points) outlining the project's background, motivation, and scope; two status reports (2.5 points each) detailing the approach, model development progress, and preliminary findings; a one-page slide with a one-minute elevator pitch (2.5 points) summarizing the project; and a comprehensive final report (40 points) with source code and detailed instructions for operation. This structure balances technical and communication skills while fostering creativity.

### *B. Student performance analysis*

The MLME course has now operated for 3 years, but it is still a relatively new concentration compared with others in the ME department at UARK. In the first year, there were about 25 students including both graduates and undergraduates. Understandably, projects by undergraduates are typically not as high-quality as projects by graduate students. This is likely due to undergraduates having insufficient research and basic programming experience. In addition, most graduate students had completed nearly all of the preferred prerequisite courses. By contrast, the undergraduates enrolled in the MLME course may have met only the minimal entry requirements, namely high school–level mathematics and physics. As a result, they were at a relative disadvantage compared to graduate students in their grasp of core concepts and in their proficiency with programming, which inevitably impacted the depth and quality of their project work and grade. These challenges remained in Fall 2022 after the trial year in Fall 2021, even though we revamped the projects for undergraduates to be more manageable and gave them some example code. At the end of Fall 2023, 14 students completed this course, which is a good sign compared to the Fall 2022 semester which had lower enrollment.

To quantify student achievement in the MLME course, the final grades of all students are recorded annually for analysis, as shown in Fig. 5. In Fall 2021, grades predominantly ranged between 60 and 85, with a median around 75. By Fall 2022, there was an improvement in performance, with grades mostly falling between 70 and 90, and several students achieving grades in the upper 80s and low 90s. This trend continued into Fall 2023, where a significant number of grades are concentrated in the high 80s and low 90s. Variance in grades was 48.49 in 2023, 42.00 in 2022, and 88.02 in 2021. In 2021, as this course was newly introduced, the instructor aimed to give students more freedom for independent exploration; therefore, no reference code was provided. While this approach encouraged creativity, it also significantly increased the difficulty of the projects, which likely contributed to the larger variance in grades that year. Inspired by cognitive load theory [48] proposed by John Sweller, which posits that working memory has a limited capacity, instructional design should avoid cognitive overload to facilitate effective knowledge acquisition and the formation of long-term memory. In 2022 and 2023, the instructor provided detailed worked examples for each project to guide students through the problem-solving process, thereby reducing extraneous cognitive load and enabling learners to devote more working memory to understanding fundamental concepts and problem-solving strategies rather than struggling with programming syntax. The impact of this adjustment was significant: as evidenced by the data, the variance in student grades decreased markedly. This outcome also supports the worked example effect [49], which suggests that providing fully guided worked solutions leads to better test performance compared to problem-solving conditions without such guidance.

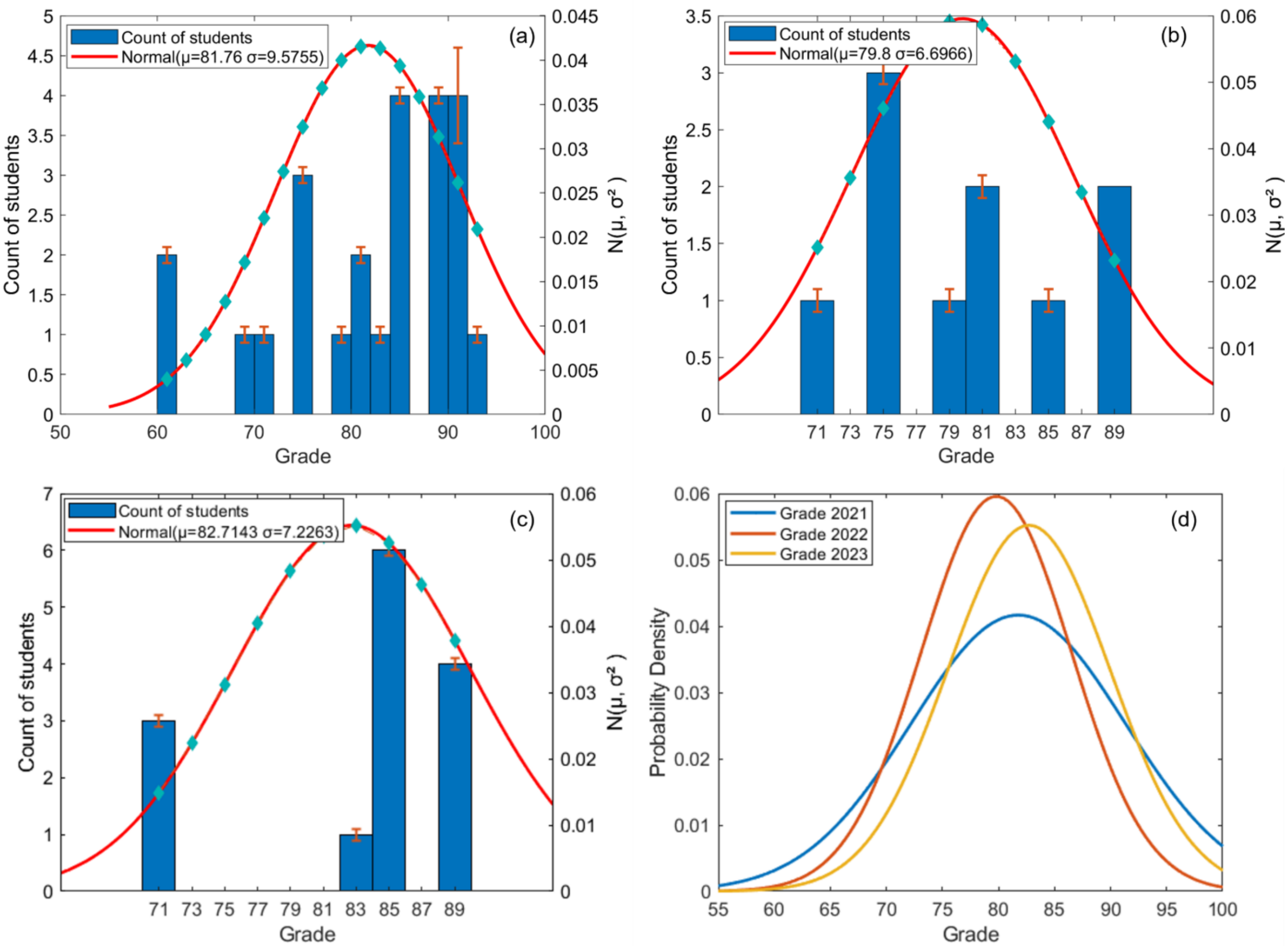

Fig. 5. Student grades for different years of the MLME course. (a) Grades of students in 2021; (b) Grades of students in 2022; (c) Grades of students in 2023; (d) Grade distributions across the three years. In Fall 2021, grades primarily ranged between 60 and 85, with a median of approximately 75, indicating a broad performance spectrum. The following year, fall 2022, saw an improvement, with grades shifting to a range of 70 to 90 and an increase in students achieving grades in the upper 80s and low 90s. This positive trend continued in Fall 2023, where a significant number of students achieved grades in the high 80s and low 90s. The variance in grades decreased over the years, from 88.02 in 2021 to 42.00 in 2022, and further to 48.49 in 2023. This reduction in grade variance suggests that providing more reference code helped students focus better on understanding key concepts and developing effective problem-solving skills. Additionally, the use of progressively complex example code appears to have facilitated improved learning outcomes and efficiency by enabling students to follow clear problem-solving steps.

*C. Examples of Student Outcomes after the MLME Course*

The class projects with student-initiated topics have led to fruitful publications. These outcomes include two parts. 1) ML model development for different engineering tasks, including manufacturing, sensing, and data analysis, as shown in Fig. 6. 2) Thermal modeling and analysis for advanced device design, including a semi-direct cooling device for high-voltage power electronic equipment and an air-cooling heat exchanger, as shown in Fig. 7.

After the Fall 2021 course, Hoskins et al. [50] developed two regression models for detecting the hole diameter in femtosecond laser processing of steel substrates, which incorporate MLP regression and GPR to measure the final diameter of pits manufactured on 304 stainless steel substrates. As shown in Fig.6(a), they processed the steel substrates using laser pulses of 0.2 seconds. Subsequently, images of the perforations were captured using an industrial camera, and datasets containing 300, 600, 900, and 1210 data points were compiled. The developed MLP and GPR were then trained and tested based on these datasets. The study finds that GPR significantly outperforms MLP in terms of detection accuracy and training speed. Hari et al. [51] developed a PCA model for non-intrusive cooling system monitoring. In this work, as shown in Fig.6(b), several boiling experiments were conducted to collect acoustic signals using both acoustic emission sensors and hydrophones as well as boiling images using a high-speed camera. Then the PCA models were applied to the three types of input data for boiling monitoring. The research results indicate that PCA was useful for two-phase cooling system monitoring based on multi-modal input. Dunlap et al. [52] developed four supervised models for assessing critical heat flux. As shown in Fig.6(c), the models included MLP, CNN, PCA-MLP, and transformers. Boiling images were collected to feed these models. This work demonstrated that all four state-of-the-art models were effective for image-based critical heat flux detection with a precision of 100%. Among them, the PCA-MLP had the highest training efficiency at 1s/epoch, and the transformer had the slowest training speed at 212s/epoch. Additionally, an unsupervised model, PCA-K-Means, was developed for comparison. The research showed that the unsupervised model can also achieve a high prediction accuracy when a large number of principal components of an image is used.

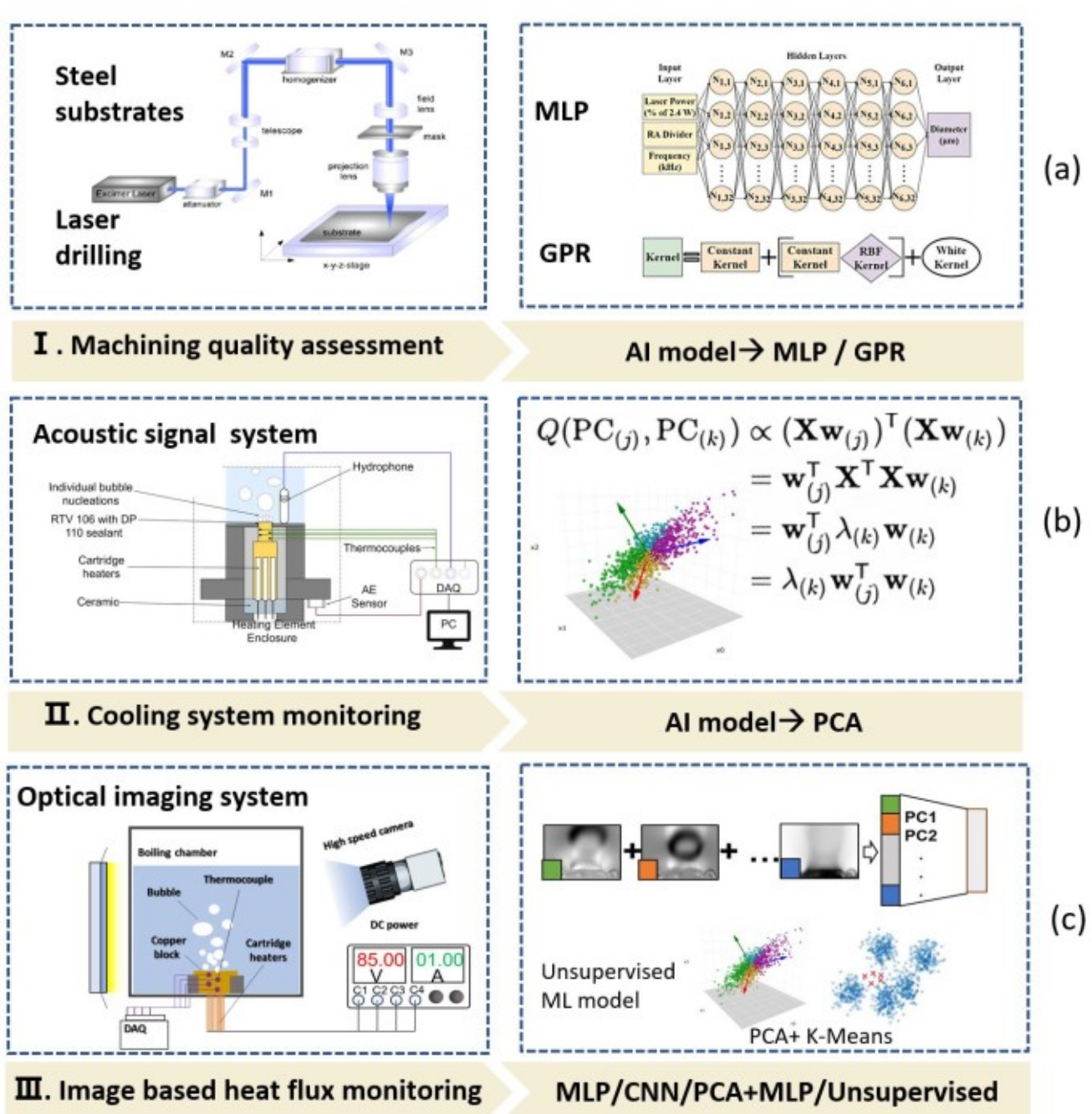

Fig. 6. Advanced outcomes on AI model development for different ME problems. (a) Outcomes on manufacturing for processed hole measurement. This outcome focuses on assessing machining quality using AI models, specifically MLP and GPR. This involves the analysis of steel substrates machined by laser drilling, where the AI models are used to predict the final hole diameter based on input parameters and data collected from the process. (b) Outcomes on sensing for two phase cooling system monitoring with multi-modal input. This outcome involves monitoring a cooling system using PCA. Acoustic signals and boiling images are collected using sensors and high-speed cameras. The PCA model processes these multi-modal inputs to monitor and diagnose the cooling system's performance, ensuring efficient operation and detecting any anomalies. (c) Outcomes on data analysis for critical heat flux warning. This outcome addresses the monitoring of heat flux using image-based methods. Multiple AI models, including MLP, CNN, and unsupervised learning techniques, are applied to analyze images captured during boiling experiments. These models help in accurately assessing the critical heat flux and understanding the heat transfer dynamics.

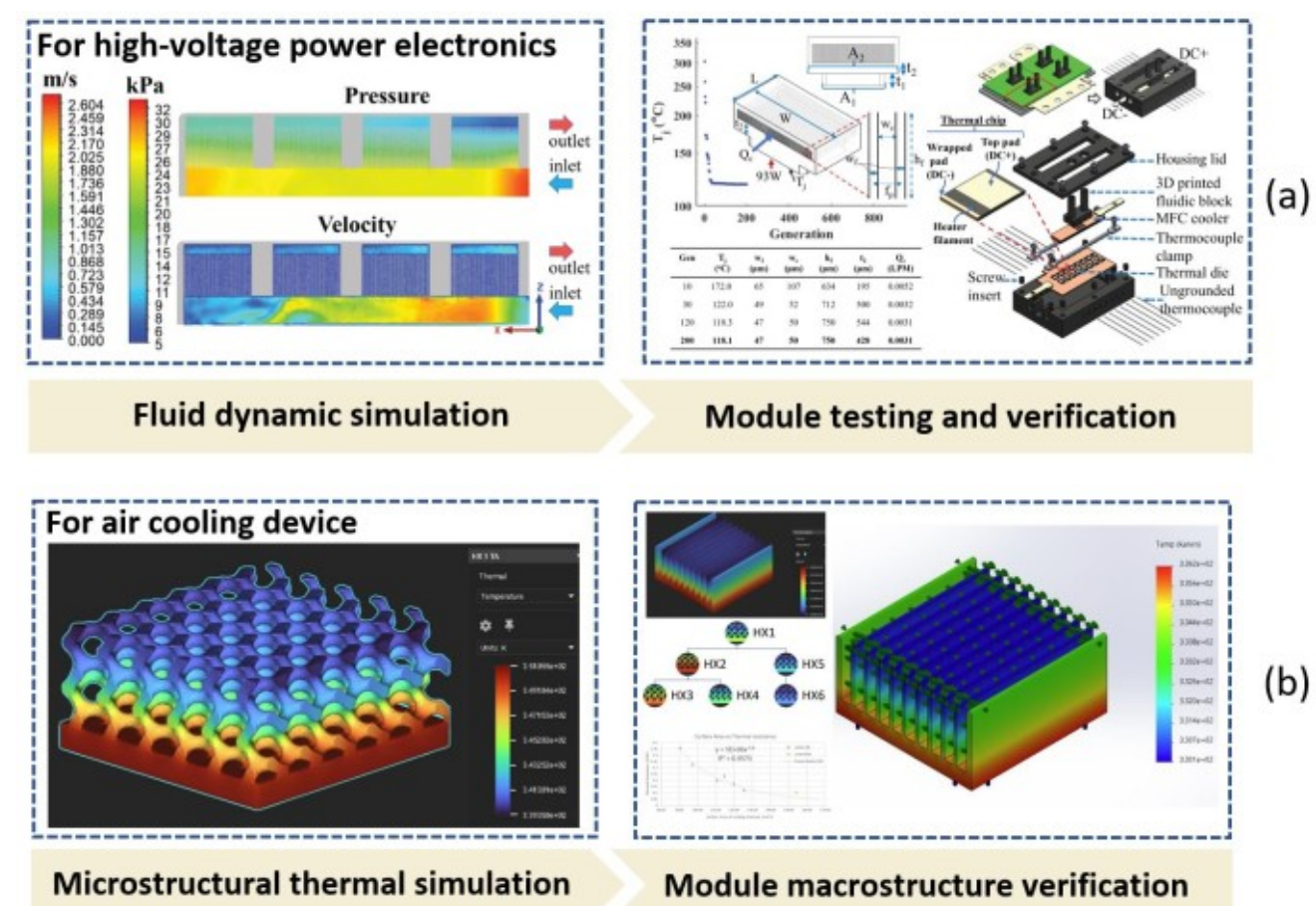

Fig. 7. Advanced outcomes for advanced device design. (a) Outcomes on fluid cooling heat exchanger design. This outcome introduced a semi-direct cooling method for silicon carbide metal-oxide-semiconductor field-effect transistors (a special type of switch used in electronics) in high-power-density half-bridge modules (components used in electronics to efficiently control and convert large amounts of electricity). This setup, equipped with a unique cooling system and connections at the top for 16 devices, is designed to manage high voltage (up to 1.7 kV) without interference. The cooling liquid (HFE7500) flows through tiny channels to keep the system cool and safe, preventing overflow and damage. (b) Outcomes on air cooling heat exchanger design. This outcome leveraged thermal knowledge from class to use nTop software for simulating and designing a new heat sink capable of high performance during steady-state convective cooling operations. Compared to standard heat exchangers, this heat sink offers similar performance and stability but with a smaller surface area.

Based on the thermal theory and AI Knowledge acquired from the MLME class, the students also built the capacity for advanced engineering designs. Iradukunda et al. [53], as shown in Fig.7(a), proposed a strategy to provide semi-direct cooling for silicon carbide metal-oxide-semiconductor field-effect transistors (a special type of switch used in electronics) in high-power-density half-bridge modules (special components used in electronics to efficiently control and convert large amounts of electricity), while also serving as a source-end interconnection (a bit like being a connecting point where electricity starts its journey). This setup has a special cooling system and connections at the top for 16 devices working together. It's designed to handle high voltage (up to 1.7 kV) without causing interference. The cooling liquid (called HFE7500) flows through tiny channels to keep things cool and safe, making sure the liquid doesn't overflow and damage the system. Based on the thermal knowledge learned in class, Miller [54] used the nTop software to simulate and design a new heat sink that can provide high performance during steady-state convective cooling operations, as shown in Fig.7(b). Compared to standard heat exchangers, the designed heat sink offers comparable performance and stability but with a smaller surface area.

## V. Conclusions

To provide ME students with solid AI-thermal knowledge and engineering problem-solving skills, an AI-integrated curriculum in thermal engineering is introduced in this paper, which is fully open source for the public. The course emphasizes project-based AI implementation capabilities in engineering. In this work, the necessity of curriculum design and the difficulties faced by students of the thermal program in ME were discussed. Then, the hierarchical and the scaffolding structure curricula were detailed. Through the four core projects and three optional projects, students can attain a comprehensive skill set and an in-depth understanding of several critical areas where AI is useful to ME and thermal engineering in particular. Based on the scores and advanced outcomes of students, the following conclusions can be drawn.

1) The project-based curricula can enhance students' understanding of the fundamental concepts, inspire the students' interest in AI tools, and improve students' performance in the coding of AI models for ME tasks.

2) The examples and scaffolded instruction on coding are important to reduce students' cognitive load, which can improve their learning outcomes and efficiency.

3) Project-based interdisciplinary courses help stimulate students' abilities to model and innovate solutions for real-world engineering problems across various fields, thereby extending the long-term impact of the curriculum on students.

4) As AI use may be different in other STEM areas, other colleagues in engineering education could take this curriculum as a starting point or as a source of new ideas for their course. And if more education materials are openly available, this could help provide a broader descriptive understanding of how AI is being used across ME, thermal, and broader fields. This can advance the field of engineering education, curriculum design, etc., more broadly.

In the future, the AI for ME curricula will be optimized further to adapt to the development of advances in AI and new challenges that face society, where ME and thermal engineering expertise is needed. In addition, more up-to-date techniques will be introduced and applied to in-class projects. We also hope to document ways in which we alter the curriculum in response to student feedback and potentially test how that might impact student engagement and student grades. Perhaps over time, we can follow up with students who have taken part in the MLME course and see if their trajectories and later outcomes in graduation, job placement, or job performance are enhanced by this new coursework. Such work can not only inform the UARK engineering curriculum, but also engineering education more broadly, and ways in which retention of engineering talent in the STEM pipeline can best be enhanced.

## AUTHOR CONTRIBUTIONS

H.H. and J.W. designed the study. C.L. and J.W. wrote the manuscript. C.D. and H.H. prepared the datasets and algorithms for the project problems. N.H. provided constructive insights into the role of AI in engineering education. All authors read and approved the final manuscript.

## ACKNOWLEDGMENTS

This study was supported by the National Science Foundation grant numbers OIA-1946391 and CBET-2323022, the University of Arkansas Chancellor's Fund for Commercialization, the Chancellor's GAP Fund, and the MathWorks Curriculum Development Support Program. This work used Bridges-2 GPU at the Pittsburgh Supercomputing Center through allocation MCH200010 from the Advanced Cyberinfrastructure Coordination Ecosystem: Services & Support (ACCESS) program, which is supported by National Science Foundation grants #2138259, #2138286, #2138307, #2137603, and #2138296. H.H. gratefully acknowledges support from the Engineering Career Connection Faculty Fellowship.

## DATA AVAILABILITY

The datasets used during the study are available in the repository MEEG-54403 on GitHub: https://github.com/hanhuark/machine-learning-for-engineers and File Exchange: https://www.mathworks.com/matlabcentral/fileexchange/166516-meeg-54403

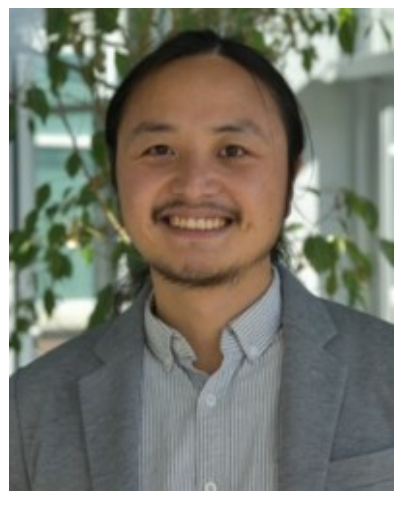

**Changgen Li** received a B.S. degree in mechatronic engineering from Xihua University, Chengdu, China, in 2018. He is a Ph.D. at the Southwest Jiaotong University, Chengdu, China, and a joint Ph.D. at Politecnico di Milano, Italy. At present, he is a research fellow at the University of Arkansas, USA. His research interests are manufacturing monitoring signal processing and machine vision inspection.

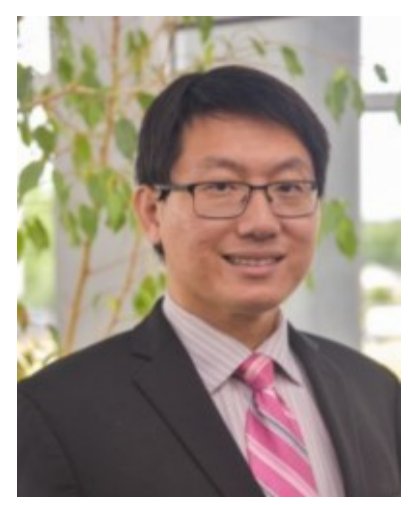

**Han Hu** (Member, IEEE) is an Associate Professor in the Department of Mechanical Engineering at the University of Arkansas. Dr. Hu received his B.S. in Theoretical and Applied Mechanics from the University of Science and Technology of China in 2011 and his Ph.D. in Mechanical Engineering from Drexel University in 2016. He was a postdoctoral researcher at the Cooling Technologies Research Center at Purdue University from 2016 to 2019 and joined the University of Arkansas in August 2019. His research centers on two-phase flows, electronic packaging, acoustic sensing, and machine learning.

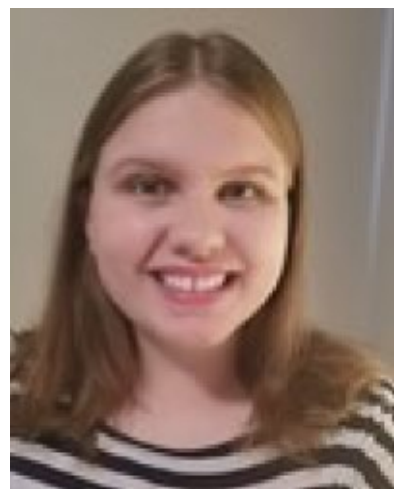

**Christy Dunlap** is a Ph.D. candidate in the Department of Mechanical Engineering at the University of Arkansas. She received her B.S. in Mechanical Engineering and B.S. in Mathematics from the University of Arkansas in 2021. Her research covers machine learning aided analysis of boiling heat transfer.

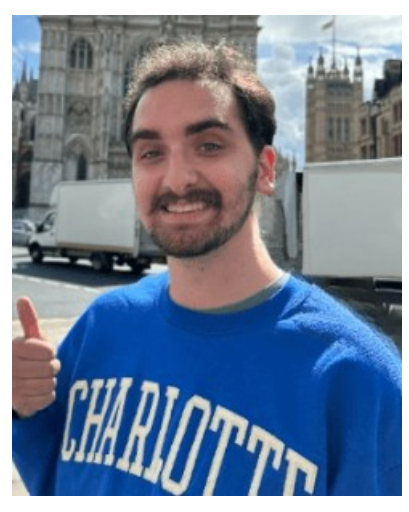

**Nathaniel House** is an undergraduate student with an interest in robotics, artificial intelligence, mechatronics, and business management. In his time on campus, he co-founded RIOT, the largest university general robotics organization, and formed it into a nonprofit organization. While running RIOT, he has since qualified himself as the engineering student body's vice president, and wants to achieve a career in robotics manufacturing/implementation, controls, artificial intelligence, or entrepreneurship of his own business.

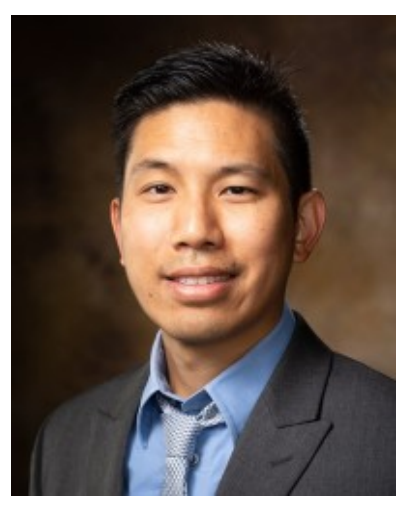

**Jonathan Wai** is Associate Professor of Education Policy and Psychology and 21st Century Chair at the University of Arkansas. His program of research is on spatial thinking and reasoning skills, talent development and K-12 education, higher education and policy, and science communication and public scholarship. His research has been funded by the U.S. Dept. of Education, he was named to the Education Week Education-Scholar Public Influence Ranking Top 200, and he is a fellow of the Association for Psychological Science. He holds a Ph.D. from Vanderbilt University and a B.A. from Claremont McKenna College.